\documentclass[lettersize,journal]{IEEEtran}
\ifCLASSINFOpdf
\else
   \usepackage[dvips]{graphicx}
\fi
\usepackage{amsmath,amsfonts}
\usepackage{amssymb}
\usepackage{diagbox}
\usepackage{algorithmic}
\usepackage{algorithm}
\usepackage{array}
\usepackage[caption=false,font=normalsize,labelfont=sf,textfont=sf]{subfig}
\usepackage{textcomp}
\usepackage{stfloats}
\usepackage{url}
\usepackage{verbatim}
\usepackage{graphicx}
\usepackage{cite}
\begin{document}

\title{A Range-Free Node Localization Method  for Anisotropic Wireless Sensor Networks with Sparse Anchors}

\author{Yong Jin {\em Member, IEEE}, Junfang Leng, Lin Zhou, {\em Member, IEEE}, Yu Jiang, Qian Wei, {\em Member, IEEE}
\thanks{This work was supported by the National Science Foundation Council of China (61976080, 61771006), the Key Research Projects of University in Henan Province, China (21A413002, 20B510001), the Programs for Science and Technology Development of Henan Province, China (212102310298, 222102210002, 222102210088), the Key Research and Development Special Project of Science and Technology Development Plan in Henan Province, China (231111212500), the Soft Science Research Program of Henan Province, China (202400410097). {\em (Corresponding author: Junfang Leng.)}}
\thanks{Y. Jin, Lin Zhou, Yu Jiang, and Qian Wei are with the School of Artificial Intelligence, Henan University, Zhengzhou 450046, China (e-mail: jy@henu.edu.cn; zhoulin@henu.edu.cn; jiang0821yu@163.com; weiqian@vip.henu.edu.cn). }
\thanks{Junfang Leng is with the School of Information Engineering, Zhengzhou College of Finance and Economics, Zhengzhou 450000, China (e-mail: ljf@henu.edu.cn). }}

\markboth{Journal of \LaTeX\ Class Files,~Vol.~14, No.~8, August~2021}%
{Shell \MakeLowercase{\textit{et al.}}: A Range-Free Node Localization Method  for Anisotropic Wireless Sensor Networks with Sparse Anchors}

\IEEEpubid{\begin{minipage}{\textwidth}\ \\[30pt] \centering
		0000--0000/00\$00.00~ \copyright ~2021 IEEE
\end{minipage}}

\maketitle

\begin{abstract}
In sensor networks characterized by irregular layouts and poor connectivity, anisotropic properties can significantly reduce the distance estimation accuracy between nodes, consequently impairing the localization precision of unknown nodes. Since distance estimation is contingent upon the multi-hop paths between anchor node pairs, assigning differential weights based on the reliability of these paths could enhance localization accuracy. To address this, we introduce an adaptive weighted method, AW-MinMax, for range-free node localization. This method involves constructing a weighted mean nodes localization model, where each multi-hop path weight is inversely proportional to the number of hops. Despite the model’s inherent non-convexity and non-differentiability, it can be reformulated into an optimization model with convex objective functions and non-convex constraints through matrix transformations. To resolve these constraints, we employ a Sequential Convex Approximation (SCA) algorithm that utilizes first-order Taylor expansion for iterative refinement. Simulation results validate that our proposed algorithm substantially improves stability and accuracy in estimating range-free node locations.
\end{abstract}

\begin{IEEEkeywords}
irregular networks, node localization, range-free, minimum maximum residual, optimization. 
\end{IEEEkeywords}

\section{Introduction}
\IEEEPARstart{W}{ireless} Sensor Networks (WSNs) are typically composed of numerous distributed sensor nodes that continuously monitor, detect, and collect data across various scenarios, such as urban surveillance, post-disaster recovery, and smart home environments \cite{b1, b2, b3}. In these contexts, accurately determining the location of each node is crucial for effective data processing \cite{b4, b5}. Although the Global Positioning System (GPS) is the predominant technology for localization, each node equipped with GPS is often impractical due to the high costs and energy consumption \cite{b6}. Consequently, a growing body of research focuses on developing localization methods that do not rely on GPS.

\subsection{Related Work}
Currently, localization schemes that do not require GPS support can be broadly classified into two categories: range-based and range-free \cite{b7}. Range-based localization algorithms, such as Received Signal Strength (RSS) \cite{b8}, Time of Arrival (ToA) \cite{b9}, and Angle of Arrival (AoA) \cite{b10}, utilize distance measurement devices within nodes to first ascertain distances between nodes, and subsequently use these distances for localization \cite{b11, b12}. In contrast, range-free methods rely on network connectivity rather than direct distance measurements to localize nodes \cite{b13, b14, b15}. Since classical range-free methods work on the assumption that nodes are uniformly distribution \cite{b16}, their performance may deteriorate in WSNs
with anisotropic factors such as irregular radio propagation and physical obstructions.

For example, by supposing nodes following Poisson distribution, the EHP model estimates inter-node distances firstly by selecting the path with the fewest hops and then employs those distance estimations to localize nodes \cite{b17}. When actual network nodes do not follow Poisson distribution, for example, in anisotropic WSNs, employing the EHP model may result in poor localization accuracy \cite{b18}. Although a two-dimensional hyperbola to model path distances can better accommodate anisotropic network characteristics \cite{b19}, this approach typically uses the mean value of multiple propagation path lengths, leading to significant localization errors. X. Yan, et al. \cite{b20} attempted to mitigate this by using a weighted average of multiple propagation path lengths to approximate inter-node distances, which somewhat improves localization accuracy. However, the method still incurs significant cumulative errors as the number of hops increases. To address this, the DV-maxHop algorithm was introduced to limit the number of hops between nodes, thus reducing cumulative errors \cite{b21}.

It is clear that these methods all employ the connectivity of the WSNs to localize nodes in scenarios where the WSNs structure is sparse—due to factors like a low density of anchor nodes or small communication radii—localization error of these methods are exacerbated with connectivity decreasing. Selecting anchor nodes strategically may be feasible to counteract the reduced accuracy caused by diminished connectivity. Although the AnSuper algorithm selects suitable anchor node pairs by supposing uniform node distribution, its accuracy declines when nodes are non-uniformly distributed \cite{b22}. While using the Geometric Dilution of Precision (GDOP) threshold can lessen the negative impacts on localization accuracy induced by the non-uniform distribution of anchor nodes, it does not address accuracy reductions due to obstacles \cite{b23}. The Reliable Anchor Pair Selection (RAPS) algorithm selects anchor node pairs based on a predefined average hop distance threshold to circumvent localization accuracy deducing induced by obstacles, achieving path distances that more closely resemble actual distances \cite{b24}. However, since it also utilizes multi-hop distances to approximate straight-line distances, the RAPS algorithm still inevitably generates cumulative errors. X. Liu, et al. \cite{b25} proposed a new algorithm with an objective function to minimize cumulative errors between unknown nodes and anchor nodes. However, in scenarios where multi-path distances between some anchor nodes are significantly greater than their real distances \cite{b26}, such as when obstacles are present, Liu’s method may experience accuracy deterioration.

\subsection{Contributions}
This paper addresses the inaccuracies arising from the disparity between the actual distances of anchor pairs and their measured distances, which are often influenced by the multi-hop path characteristics of the anchor pair in anisotropic WSNs. To enhance node localization accuracy, our approach involves assigning different weights to multi-hop paths based on their distance estimation accuracy. Building on this concept, we introduce a weighted mean nodes localization model and its corresponding solver. The primary contributions of our work include:
\subsubsection{} We observe that a greater hop count between the given anchor pair typically leads to larger errors in distance estimation. To counter this, we have developed a weighted mean nodes localization model where each multi-hop path is assigned a weight inversely proportional to its hop count, aiming to mitigate the localization accuracy loss caused by multi-hop paths with great hop count.
\subsubsection{} To handle the proposed model's inherent non-convexity and non-differentiability, we utilize matrix transformations to reformulate the proposed model into an optimization problem with a convex objective function and non-convex constraints. 
\subsubsection{} To address the non-convex constraints, we first apply a first-order Taylor expansion to tighten them. Subsequently, we introduce a Successive Convex Approximation (SCA) algorithm designed to solve the proposed model iteratively. This algorithm enhances the feasibility and accuracy of solving the non-convex optimization problem, progressively improving the model's performance through iterations.
\subsection{ Paper Organization}
The remainder of this paper is structured as follows: Section \ref{section2} outlines the system architecture and discusses the localization challenges within irregular networks. Section \ref{section3} provides a detailed introduction to the proposed algorithm, including its pseudo-code, computational complexity, and additional relevant aspects. Section \ref{section4} demonstrates the feasibility and effectiveness of the proposed algorithm through simulation experiments. Finally, Section \ref{section5} presents conclusions drawn from the research findings.

\begin{figure}[htbp]
\includegraphics[width=3.3in]{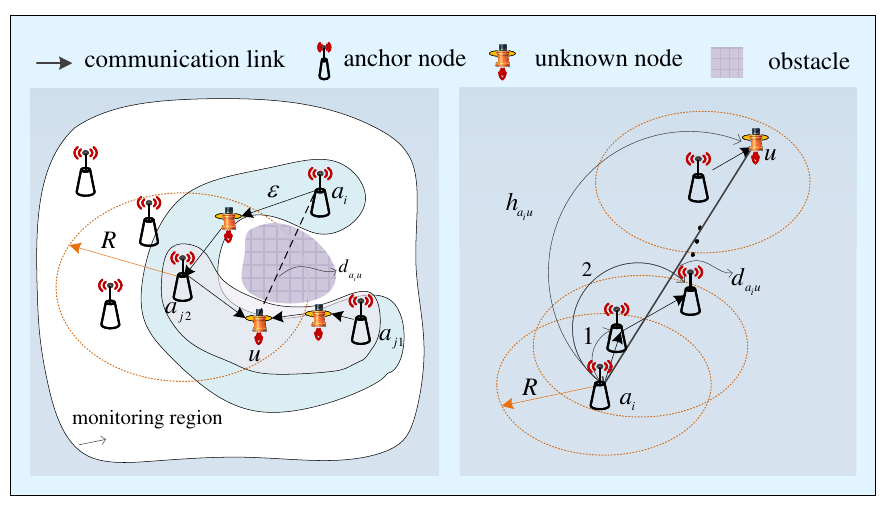}
\caption{WSNs model in underwater.\label{fig1}}
\end{figure}
\section{System model and problem description}
\label{section2}
The monitoring of Wireless Sensor Networks (WSNs) in underwater activities, such as frogman operations, wreck exploration, and Autonomous Underwater Vehicle (AUV) navigation, is critically important. For simplicity, the two-dimensional WSNs model is represented as an undirected graph ${\cal G} = \left( {{\cal N},{\cal E}} \right)$ as illustrated in Fig.\ref{fig1}. ${\cal E}$ is a set of edges between the two nodes. ${\cal N} \buildrel \Delta \over = {{\cal N}_a} \cup {{\cal N}_u}$ is composed of anchor node set ${{\cal N}_a}$ and unknown node set ${{\cal N}_u}$. The cardinal of ${\cal N}$ and ${{\cal N}_a}$ are represented by $n \buildrel \Delta \over = \left| {\cal N} \right|$ and $m \buildrel \Delta \over = \left| {{{\cal N}_a}} \right|$, respectively. ${\textbf{\emph{x}}_i} = {\left[ {{x_i},{y_i}} \right]^{\rm{T}}}$ is defined as the location coordinate of a node $i\left( {i \in {\cal N}} \right)$ in the WSNs. 

Utilizing the multilateration localization principle based on least squares \cite{b27}, the coordinate estimation of an unknown node $u$ is obtained. The DV-Hop algorithm, which typically employs a linear model, often results in significant localization errors due to its simplistic approach. To improve the localization accuracy of an unknown node with a location coordinate $\textbf{\emph{x}}_{u}$, a nonlinear model utilizing the min-max error criterion is proposed \cite{b29}. The optimization problem is formulated as:
\begin{equation}
\label{eq1}
	\underset{{\textbf{\emph{x}}_{u}}}{\mathop{\min }}\,\underset{a}{\mathop{\max }}\,\left\{ {\left| {{\left\| {\textbf{\emph{x}}_{u}}-{\textbf{\emph{x}}_{{{a}_{i}}}} \right\|}_{2}}-{{{\hat{d}}}_{{{a}_{i}}u}} \right| }\right\}
\end{equation} 
where ${\left\| \cdot \right\|_{2}}$ denotes the Euclidean norm, $a\triangleq \left\{ {{a}_{i}}|i=1,\ldots, m \right\}$ represents the set of anchor nodes, and ${\textbf{\emph{x}}_{{{a}_{i}}}}$ specifies the position of the $i$-th anchor node. This approach aims to refine the accuracy by minimizing the maximum estimated error between the actual and the calculated distances to the anchor nodes.

Assuming that all nodes use the DV-Hop strategy \cite{b27}, namely (\ref{eq2}), to estimate the distance ${{\hat{d}}_{{{a}_{i}}u}}$ between the unknown node $u$ and the anchor node ${{a}_{i}}$. 
\begin{equation}
\label{eq2}
{{\hat{d}}_{{{a}_{i}}u}}={{\bar{d}}_{{{a}_{i}}}}\times {{h}_{{{a}_{i}}u}}
\end{equation}
where ${{h}_{{{a}_{i}}u}}$ denotes hop count between anchor node ${a}_{i}$ and unknown node $u$, ${{\bar{d}}_{{{a}_{i}}}}$ described in (\ref{eq3}) is average hop distance of anchor node ${{a}_{i}}$.
\begin{equation}
\label{eq3}
{{\bar{d}}_{{{a}_{i}}}}\text{=}\frac{\sum\limits_{j\ne i}{{{d}_{{{a}_{i}}{{a}_{j}}}}}}{\sum\limits_{j\ne i}{{{h}_{{{a}_{i}}{{a}_{j}}}}}},\text{ }i,j=1,\cdots ,m.
\end{equation}
where ${{d}_{{{a}_{i}}{{a}_{j}}}}$ , ${{h}_{{{a}_{i}}{{a}_{j}}}}$ denotes distance and hops between
anchor node ${a}_{i}$ and ${a}_{j}$, respectively.

Examining model (1), it becomes evident that its performance is heavily dependent on the accuracy of ${{\hat{d}}_{{{a}_{i}}u}}$, which is directly tied to the effectiveness of the DV-Hop algorithm. While the DV-Hop algorithm may demonstrate high accuracy in WSNs with uniformly distributed nodes, various factors, such as physical obstacles and irregular radio propagation, can result in a non-uniform distribution of nodes. This non-uniformity can significantly impair the performance of the DV-Hop algorithm (as detailed later in the text). Consequently, directly applying model (1) to localize common nodes under such conditions is not advisable due to the potential for severe accuracy degradation.

\begin{figure}[htbp]
\centering
\includegraphics[width=3.3in]{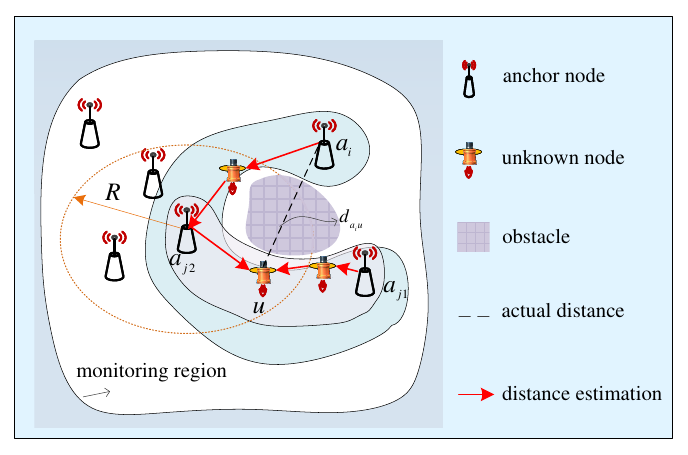}
\caption{Network with obstacle.}
\label{fig2}
\end{figure}

The accuracy degradation of the DV-Hop algorithm, induced by obstacles, is illustrated in Fig.\ref{fig2}. This network node distribution shows the deployment of nodes in a real scenario where the center contains a circular obstacle. The sensor nodes are non-uniformly distributed within the monitoring area because nodes cannot be deployed in the monitoring area where obstacles are present. One can find that obstacles may block signal propagation. Thus, distance estimation ${{\hat{d}}_{{{a}_{i}}u}}$ employed by using the DV-Hop algorithm may deviate its real value ${{d}_{{{a}_{i}}u}}$.

For instance, in scenarios where each node's communication radius is set to be $R$, a direct communication link exists between two nodes if their distance is less than $R$. However, obstacles can disrupt this direct connectivity, forcing the communication path between nodes to detour. For example, the unknown node $u$ is supposed to connect with the anchor nodes ${{a}_{i}}$, ${{a}_{j1}}$, and ${{a}_{j2}}$. It is clear that three hops path distances between ${{a}_{j1}}$ and ${{a}_{j2}}$ may closely approximate the actual distances, while the presence of an obstacle disrupts signal transmission between ${{a}_{i}}$ and ${{a}_{j1}}$, leading to five hops between anchor nodes ${{a}_{i}}$${{a}_{j1}}$ apparently different with actual distance between ${{a}_{i}}$ and ${{a}_{j1}}$. 

Following the above description, we may deduce that reasonably weighting different multi-hop paths, namely giving a high confidence level to those multi-hop paths with high estimation accuracy, may enhance the localization accuracy of unknown nodes. Inspired by this idea, we revise the model (1) and propose a novel localization scheme in the next section.
\begin{figure}[htbp]
\centering
\includegraphics[width=3.3in]{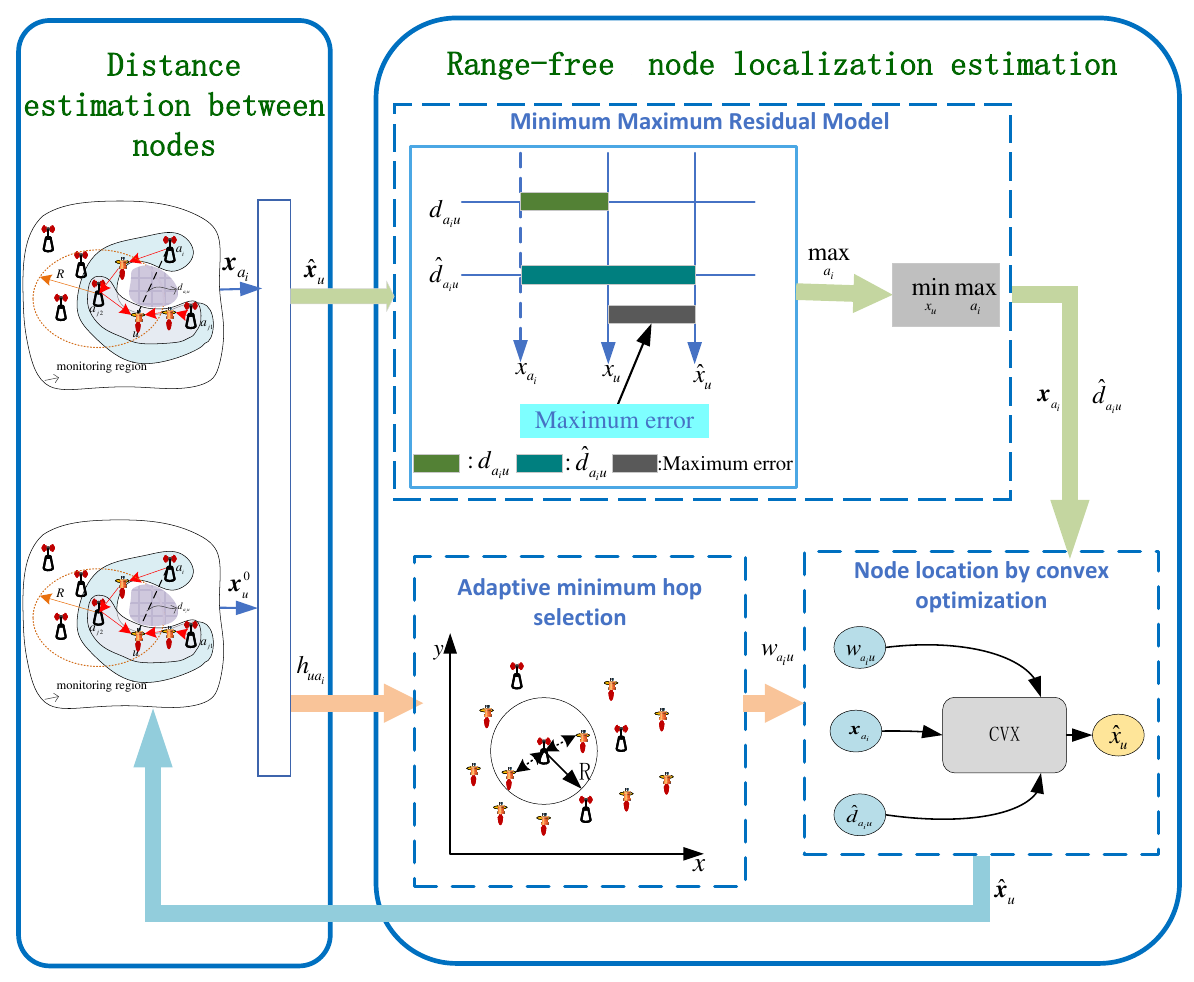}
\caption{AW-MinMax algorithm.}
\label{fig3}
\end{figure}

\section{Proposed localization method}
\label{section3}
The proposed a novel localization scheme, named Adaptive Weighted Minimum Maximum Residual (AW-MinMax), which combines novel distance estimation between nodes and weight multi-hop path to tackle the challenges of low localization accuracy prevalent in irregular WSNs.

By strategically classifying anchor node pairs according to their confidence level of distance estimation. Subsequently, we first refine the distance estimation process between nodes. Then, we bolster range-free node localization by assigning different weights according to counts of multi-hop paths and estimation accuracy of the distance between nodes, as illustrated in Fig.\ref{fig3}. 
\subsection{Anchor Pairs Classify and  Distance Estimation Between Nodes}
This section introduces a novel method which can adaptive selects scheme of distance estimation between nodes according to the type of anchor node pairs.
\begin{figure}[htbp]
\centering
\includegraphics[width=3.3in]{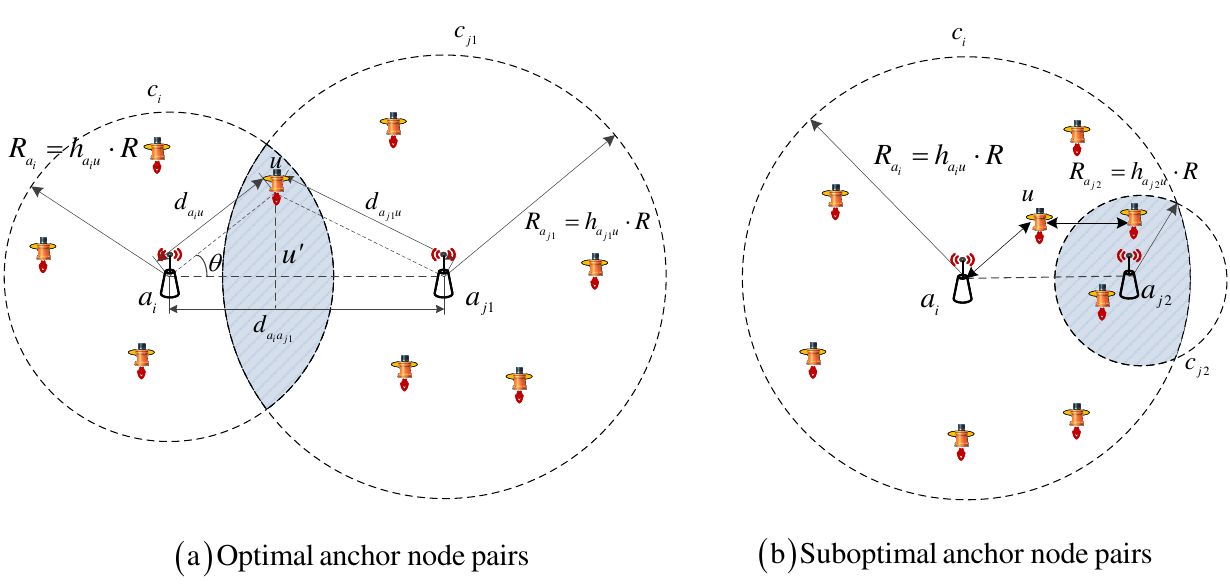}
\caption{Anchor node pairs.}
\label{fig4}
\end{figure}
\subsubsection{Novel anchor node pairs types}
As depicted in Fig.\ref{fig4}, according to estimation accuracy of distance between the nodes, anchor pairs can be categorized into three types as follows \cite{b25}.

\paragraph{Optimal anchor node pairs}

According to the distribution of anchor
nodes (as illustrated in Fig.\ref{fig4}(a)), the
anchor node pairs, the anchor nodes pairs which are composed of anchor node ${{a}_{i}}$ and ${{a}_{j1}}$ are defined as optimal anchor node pairs with respect to unknown node u if these nodes pair follows (\ref{eq4}).
\begin{equation}
\label{eq4}
  \begin{aligned}&condition\text{1-1}:d_{_{a_ia_{j_1}}}>R\cdot h_{_{a_iu}}\quad{and}\quad d_{_{a_ia_{j_1}}}>R\cdot h_{_{a_{j_1}}u}\\&condition\text{ 1-2}:-1\leq\frac{\left(R\cdot h_{_{a_iu}}\right)^2+d_{_{a_ia_{j_1}}}^2-\left(R\cdot h_{_{a_{j_1}u}}\right)^2}{2R\cdot h_{_{a_iu}}\cdot d_{_{a_ia_{j_1}}}}\leq1\end{aligned}
\end{equation}

\paragraph{Suboptimal anchor node pairs}

Similarly to the optimal anchor node pairs, suboptimal anchor node pairs with respect to unknown node u are defined as (\ref{eq5}).

\begin{equation}
\label{eq5}
\begin{aligned}&condition\text{ 2-1}:d_{_{a_ia_{j2}}}<R\cdot h_{_{a_iu}}\quad and\quad d_{_{a_ia_{j2}}}>R\cdot h_{_{a_{j2}u}}\\&condition\text{ 2-2}:-1\leq\frac{\left(R\cdot h_{_{a_iu}}\right)^2+d_{_{a_ia_{j_2}}}^2-\left(R\cdot h_{_{a_{j_2}u}}\right)^2}{2R\cdot h_{_{a_iu}}\cdot d_{_{a_ia_{j_2}}}}\leq1\end{aligned}
\end{equation}

\paragraph{Unavailable anchor node pairs}

Unavailable anchor node pairs include two cases. The one is the distance between the anchor pairs too far, namely the maximum coverage area of each node in these anchor pairs non-overlap at all. The other one is the distance between the anchor pairs too near, namely the unknown nodes on the same side of the anchor pairs. Consequently, these pairs cannot provide reliable information for localization, rendering the conditions for optimal and suboptimal anchor pairs inapplicable. Thus, estimating the distance between unknown nodes and anchor nodes becomes unfeasible.
\subsubsection{Distance estimation between nodes according to anchor pairs category}
Once the type of anchor node pairs is determined, one can employ different schemes to estimate distances between nodes according to those types. 

\paragraph{Distance estimation of optimal anchor node pairs}

Given the optimal anchor node pairs, similar to the work of \cite{b25}, the distance between an unknown node $u$ and an anchor node ${{a}_{i}}$ is estimated by using (6):
\begin{equation}
\label{eq6}
	{{\hat{d}}_{{{a}_{i}}u}}=\int_{0}^{{{\theta }_{o{{a}_{i}}{{a}_{j1}}}}}{{{p}_{\theta _{{{a}_{i}}}^{u}}}}\left( \theta  \right)\frac{{{{\bar{d}}}_{{{a}_{i}}}}\cdot {{h}_{{{a}_{i}}u}}}{\cos \theta }d\theta 
\end{equation}
where, $\theta _{{{a}_{i}}}^{u}$ represents the angle between the anchor node pair ${{A}_{{{a}_{i}}{{a}_{j1}}}}$
and the unknown node $u$. ${{\bar{d}}_{{{a}_{i}}}}$ denotes the average hop distance. ${{p}_{\theta _{{{a}_{i}}}^{u}}}\left( \theta  \right)$ is the probability density of $\theta _{{{a}_{i}}}^{u}$. Notably, we proposed a novel method that is different from the work of \cite{b25} to obtain ${{p}_{\theta _{{{a}_{i}}}^{u}}}\left( \theta  \right)$ (details in AppendixI). 

\paragraph{Distance estimation of suboptimal anchor node pairs}

Due to geometric structure of suboptimal anchor node pairs is different with that of optimal node pairs, directly employing (6) to estimate distance between the suboptimal anchor node pairs is infeasible. We proposed the other distance estimation method, which is written in (7):
\begin{equation}
\label{eq7}
	{{\hat{d}}_{{{a}_{i}}u}}=\frac{{{d}_{{{a}_{i}}{{a}_{j2}}}}}{{{h}_{{{a}_{i}}{{a}_{j2}}}}}\cdot {{h}_{{{a}_{i}}u}}
\end{equation}
where ${{h}_{{{a}_{i}}u}}$ and ${{h}_{{{a}_{i}}{{a}_{j2}}}}$ represent the hop from the anchor node ${{a}_{i}}$ to the unknown node $u$ and the anchor node ${{a}_{j2}}$, respectively. ${{d}_{{{a}_{i}}{{a}_{j2}}}}$ represents the multi-hop distance between the anchor node ${{a}_{i}}$ and ${{a}_{j2}}$.

\subsection{Range-free Node Localization Estimation}
Once the ${{\hat{d}}_{{{a}_{i}}u}}$ to be obtained, we can revise range-free node localization by assigning differentiated weights according to the counts of the multi-hop paths since it experiences negative correlation with accuracy of ${{\hat{d}}_{{{a}_{i}}u}}$. As illustrated in Fig.\ref{fig3}, the process of range-free node localization estimation encompasses three primary components: constructing a minimum and maximum residual model, implementing adaptive minimum hop selection, and determining node locations through SCA. Here's a detailed breakdown of each component:
\subsubsection{ Minimum Maximum residual model}According to the above descriptions,  we can conclude that different anchor node pairs across one unknown node mean different multi-hop paths, while different multi-hop paths may lead to different localization accuracies of unknown nodes. Generally, fewer counts of multi-hop paths mean better unknown node localization accuracy. Inspired by this intuition, we can revise the optimization problem (1) to the following description.
\begin{equation}
\label{eq8}
	\underset{{\textbf{\emph{x}}_{u}}}{\mathop{\min }}\,\underset{a}{\mathop{\max }}\,\left\{ {{w}_{{{a}_{i}}u}}\left| {{\left\| {\textbf{\emph{x}}_{u}}-{\textbf{\emph{x}}_{{{a}_{i}}}} \right\|}_{2}}-{{{\hat{d}}}_{{{a}_{i}}u}} \right| \right\}
\end{equation} 
where ${{w}_{{{a}_{i}}u}}$ is the adaptive weighting which is determined by count of multi-hop paths. 
\subsubsection{ Adaptive minimum hop selection}
Observing (8), one can find that it is an extension version of the optimization problem (1). Since in uniformly WSN, multi-hop paths are almost irrelevant with a localization accuracy of unknown nodes, the weight related to the first item of (1) is set to be 1. While in irregular WSNs, as the above descriptions, different multi-hop paths across one unknown node mean different localization accuracy; thus, it is reasonable to give large weight to those multi-hop paths with higher localization accuracy. Now, we will explore how to choose rational weight ${{w}_{{{a}_{i}}u}}$ to enhance localization accuracy here. Based on the above discussion, we can conclude that less minimum hop between the anchor node ${{a}_{i}}$ and the unknown node $u$ means high localization accuracy; moreover, the estimation accuracy of ${{\bar{d}}_{{{a}_{i}}}}$ will determine the localization accuracy of unknown node too. Thus, we design the weight related to different multi-hop paths as follow:
\begin{equation}
\label{eq9}
	{{w}_{{{a}_{i}}u}}=\frac{1}{{{h}_{{{a}_{i}}u}}^{\Delta \sigma }}
\end{equation} 
where ${{h}_{{{a}_{i}}u}}$ denotes the minimum hop count between the anchor node ${{a}_{i}}$ and the unknown node $u$ , and the exponent $\Delta \sigma $ is defined as:
\begin{equation}
\label{eq10}
	\Delta \sigma ={{\bar{d}}_{_{{{a}_{i}}}}}\times ({{h}_{{{a}_{i}}u}}+{{h}_{{{a}_{j}}u}})-{{d}_{{{a}_{i}}{{a}_{j}}}}
\end{equation}
where $\Delta \sigma $ denotes the difference between the multi-hop distance and the actual distance between anchor nodes ${{a}_{i}}$ and ${{a}_{j}}$.

Overall, ${{w}_{{{a}_{i}}u}}$ is inversely proportional to ${{h}_{{{a}_{i}}u}}$. This means that the larger the minimum hop count ${{h}_{{{a}_{i}}u}}$, the greater the potential error in the distance estimation ${{\hat{d}}_{{{a}_{i}}u}}$, thus the smaller ${{w}_{{{a}_{i}}u}}$ can be used to suppress the larger error concealed in the first item of the (8). Moreover, supposing one multi-hop path with large difference $\Delta \sigma $, which means using those multi-hop paths to localize unknown nods may induce heavy error. Thus, we can obtain little ${{w}_{{{a}_{i}}u}}$ by employing (9) in this case, which indicates that impact on localization accuracy induced by multi-hop paths with large $\Delta \sigma $ may be weakened.
\subsubsection{Node location by using SCA method}
Clearly, the adaptive weighted minimum maximum residual optimization problem expressed in (\ref{eq8}) is characterized by its non-differentiability and non-convex nature \cite{b29}. To address this, we employ the Chebyshev approximation by introducing a transformation variable 
$t=\underset{{{a}_{i}}}{\mathop{\max }}\,\left\{ {{w}_{{{a}_{i}}u}}\left| {{\left\| {\textbf{\emph{x}}_{u}}-{\textbf{\emph{x}}_{{{a}_{i}}}} \right\|}_{2}}-{{{\hat{d}}}_{{{a}_{i}}u}} \right|,\text{ }i=1,\cdots ,m \right\}$ into (\ref{eq8}). By doing so, we can reformulate the original problem of the new model into a more tractable optimization problem which is described as (\ref{eq11}).
\begin{subequations}
  \begin{eqnarray}
  \label{eq11}
  &\mathop {\mathrm{min }}\limits_{{\textbf{\emph{x}}}_u, t} & {t} \\
  &\mathrm{s.t.} & w_{a_i u} \text{[} \left\| {\textbf{\emph{x}}}_u - {\textbf{\emph{x}}}_{a_i} \right\|_2 - {\hat d}_{a_i u} \text{]} \ge -t, \notag\\
  & & i = 1, 2,\cdots ,m.\\
  & & w_{a_i u} \text{[}\left\| {\textbf{\emph{x}}}_u - {\textbf{\emph{x}}}_{a_i} \right\|_2 - {\hat d}_{a_i u} \text{]}\le t , \notag\\
  & & i = 1, 2,\cdots ,m.
  \end{eqnarray}
\end{subequations}

One can find that the objective function (11a) and the inequalities constraint (11c) are convex, while the inequalities constraint (11b) exhibits nonconvexity. We can tighten it by using a first-order Taylor approximation, which is described in (\ref{eq12}):
\begin{equation}
\label{eq12}
    \begin{array}{l}
    {{w}_{{{a}_{i}}u}}{{\text{[}\left\| {{\textbf{\emph{x}}}_{u}^{0}}-{{\textbf{\emph{x}}}}_{{{a}_{i}}} \right\|}_{2}}-{{\hat d}_{{{a}_{i}}u}}+\frac{{{\left( {{{\textbf{\emph{x}}}}_{u}^{0}}-{\textbf{\emph{x}}_{{{a}_{i}}}} \right)}^{\rm{T}}}}{{{\left\| {{\textbf{\emph{x}}}_u^0}-{\textbf{\emph{x}}_{{{a}_{i}}}} \right\|}_{2}}}\left( {{\textbf{\emph{x}}_{u}}-{{\textbf{\emph{x}}}_u^0}} \right)\text{]}\ge -t, \\
    i = 1, 2,\cdots ,m.
    \end{array}
\end{equation}
where ${\textbf{\emph{x}}_{u}^{0}}$ is initial location coordinate of unknown node $u$.

Correspondingly, the optimization problem (11) can be transformed into a structured optimization problem with the following constraints:
\begin{equation}
\label{eq13}
\begin{array}{l}
\mathop {\mathrm{min }}\limits_{ {t},\ {{\textbf{\emph{x}}}_u}}\,\, {t}\\
\mathrm{s.t.}  \,\,{w}_{{{a}_{i}}u}{\text{[}\left\| {{{\textbf{\emph{x}}}_u^0}- {{\textbf{\emph{x}}}_a}} \right\|_2} - {{\hat d}_{{{a}_{i}}u}} + \frac{{{{\left( {{{\textbf{\emph{x}}}_u^0} - {{\textbf{\emph{x}}}_{{a}_{i}}}} \right)}^{\rm{T}}}}}{{{{\left\| {{\textbf{\emph{x}}}_u^0 - {{\textbf{\emph{x}}}_{{a}_{i}}}} \right\|}_2}}}\left( {{{\textbf{\emph{x}}}_u} - {{\textbf{\emph{x}}}_u^0}} \right) \text{]}\ge  - {t}, \\
\quad\quad i = 1, 2,\cdots ,m.\\
\quad\quad {w}_{{{a}_{i}}u}{\text{[}\left\| {{{\textbf{\emph{x}}}_u}- {{\textbf{\emph{x}}}_{{a}_{i}}}} \right\|_2} - {{\hat d}_{{{a}_{i}}u}}\text{]} \le {t}, i = 1, 2,\cdots ,m.
\end{array}
\end{equation}

According to the above analysis, the non-convex and non-differentiable problem (\ref{eq8}) has been tightened to a convex optimization problem (\ref{eq13}).
 By utilizing an iterative approach, we can efficiently converge to the optimal solution ${\textbf{\emph{x}}_{u}^{k}}$ that satisfies a specified accuracy threshold $\mathsf{\epsilon}$. Therefore, given an initial location, unknown node location ${\textbf{\emph{x}}_{u}}$ can be easily obtained by using the gradient descent method. More details related to the proposed algorithm are shown in Algorithm \ref{alg:alg2}.
 \begin{algorithm}[H]
\caption{AW-MinMax Algorithm}\label{alg:alg2}
\begin{algorithmic}
\STATE 
\STATE {\textsc{I}nput}: Communication range $R$; Anchor node location \\
\quad\quad sets $\left\{{\textbf{\emph{x}}}_{a_i} \right\}$; Accuracy threshold $\epsilon$;
\STATE {\textsc{O}utput}: Unknown node location ${\textbf{\emph{x}}}_u$;
\STATE {1.} Determine the type of anchor node pair according to (\ref{eq4}) \\
\quad and (\ref{eq5}); \\
\STATE {2.} For optimal anchor node pairs, use (\ref{eq6}) to obtain $\hat{d}_{{a_{i}}u}$, \\
\quad for suboptimal anchor node pairs, use (\ref{eq7}) to obtain $\hat{d}_{{a_{i}}u}$; \\
\STATE {3.} Calculate the weight factor $w_{{a_{i}}u}$ using (\ref{eq9}); \\
\STATE {4.} Give initial location ${{\textbf{\emph{x}}}_u^0}$ of unknown node; \\
\STATE {5.} \textbf{Repeat:} \\
\STATE {6.} \quad Obtain optimal solution ${{\textbf{\emph{x}}}_u^{k}}$ through solving optimiz-\\
 \quad\quad ation problem (13);\\
\STATE {7.} \quad ${{\textbf{\emph{x}}}_u^{k-1} \leftarrow {\textbf{\emph{x}}}_u^k}$; \\
\STATE {8.} \textbf{Until:} ${\left\| {\textbf{\emph{x}}_u^{k} - \textbf{\emph{x}}_u^{k-1}} \right\|_2} < \epsilon $; \\
\end{algorithmic}
\label{alg2}
\end{algorithm}

\section{Complexity analysis}
\label{section4}
This section delves into the computational complexity of the proposed AW-MinMax algorithm, which primarily hinges on two factors: the number of iterations and the computational cost per iteration.

1) Number of Iterations:
From the standpoint of iterations, the computational complexity involved in calculating the average hop distance for anchor node pairs, as shown in (\ref{eq3}), is $\mathcal{O}\left( m\cdot \left( m-1 \right)/2 \right)$. Additionally, the complexity for estimating the distance between nodes is $\mathcal{O}\left( m \right)$.

2) Computational Cost Per Iteration:
In terms of computational costs per iteration, the optimization problem, as defined in our algorithm, comprises three variable elements, $m$ linear matrix inequality (LMI) constraints and $m$ second-order cone (SOC) inequality constraints. The cost required to achieve the accuracy threshold $\epsilon $ through the necessary number of iterations is approximated as 
$\sqrt{4m}\ln \left( 1/\epsilon  \right)$. Consequently, the total computational complexity per unknown node can be expressed as:
$\mathcal{O}\left( {{m}^{2}}+m/2+\sqrt{4m}\ln \left( 1/\epsilon  \right) \right)$ Upon examining the computational complexities of other comparison algorithms  DV-Hop, Hyperbolic, AAML, and LAPCD, it is evident from Table \ref{tab1} that, compared to the DV-Hop, Hyperbolic, and AAML algorithms, the complexity of the proposed AW-MinMax algorithm is reduced by two orders of magnitude. However, it is slightly higher than that of the LAPCD algorithm.
This analysis highlights the efficiency of the AW-MinMax algorithm in managing computational resources while maintaining a competitive edge in performance compared to existing methodologies.

\begin{table}[!t]
\renewcommand{\arraystretch}{1.3}
\caption{Computational complexity\label{tab1}}
\setlength{\tabcolsep}{3pt}
\begin{tabular}{p{50pt}p{115pt}p{65pt}}
\hline
Algorithm& 
Complexity in general& 
Complexity in the simulation scenes \\
\hline
DV-Hop\cite{b28}& 
$\mathcal{O}\left( {{m}^{3}} \right)$& 
$3.28\times {{10}^{7}}$ \\
Hyperbolic\cite{b22}& 
$\mathcal{O}\left( {{m}^{3}} \right)$& 
$3.28\times {{10}^{7}}$ \\
AAML\cite{b23}& 
$\mathcal{O}\left( {{m}^{2}}+{{c}^{2}}m+{{c}^{3}} \right)$& 
$1.07\times {{10}^{7}}$ \\
LAPCD\cite{b25}& 
$\mathcal{O}\left( {{m}^{2}}+m/2+Pops\cdot I \right)$& 
$1.04\times {{10}^{5}}$ \\
AW-MinMax& 
$\mathcal{O}\left( {{m}^{2}}+m/2+\sqrt{4m}\ln \left( 1/\epsilon  \right) \right)$& 
$1.03\times {{10}^{5}}$\\
\hline
\multicolumn{3}{p{251pt}}{Note: In Table \ref{tab1}, $c$ denotes the total number of features/variables, $Pops$ and $I$ denote the population size and number of iterations, respectively. }\\
\end{tabular}
\end{table}

\section{Simulation results and analysis}
\label{section5}
In this section, we outline the experimental scenario settings, algorithm selection, and performance metrics for algorithm evaluation and provide comparative results to validate the effectiveness of the proposed AW-MinMax algorithm. This comprehensive overview aims to demonstrate the algorithm's robustness and accuracy in various testing environments.
\subsection{Scenario Settings And Algorithm selection}
Experiment scenario is designed according to the Fig.\ref{fig2} and parameters related to this scenario are listed in Table \ref{tab2}.
\begin{table}[!t]
\renewcommand{\arraystretch}{1.3}
\caption{Simulation Parameters\label{tab2}}
\setlength{\tabcolsep}{3pt}
\begin{tabular}{p{135pt}p{49pt}p{46pt}}
\hline
Parameter& 
Symbol& 
Value \\
\hline
Area of simulation$({{m}^{2}})$& 
$L\times L$& 
100$\times$100 \\
Total number of nodes& 
$n$& 
150 \\
Number of anchor nodes& 
$m$& 
30 \\
Number of unknown nodes& 
$n-m$& 
120 \\
Communication radius $\left( m \right)$& 
$R$& 
20\\
Degree of Irregularity& 
$DoI$& 
0.02\\
Threshold& 
$\epsilon $& 
${10}^{-3}$\\
Simulation times& 
${{M}_{c}}$& 
100\\
\hline
\end{tabular}
\end{table}

To benchmark the performance of our proposed AW-MinMax algorithm, we have selected traditional range-free node localization algorithms including DV-Hop, Hyperbolic, AAML, and LAPCD (details in Table \ref{tab3}) for comparison.

\begin{table}[!t]
\renewcommand{\arraystretch}{1.3}
\caption{Comparison algorithms\label{tab3}}
\setlength{\tabcolsep}{3pt}
\begin{tabular}{p{95pt}p{40pt}p{40pt}p{40pt}}
\hline
\diagbox {Algorithm}{Performance}& 
Precision&
Complexity&
Stability \\
\hline
DV-Hop\cite{b28}& 
Low& 
High& 
Low \\
Hyperbolic\cite{b22}& 
Medium& 
High& 
Low \\
AAML\cite{b23}& 
Medium& 
Medium& 
Medium \\
LAPCD\cite{b25}& 
High& 
Low& 
Medium \\
AW-MinMax& 
High&
Low& 
High \\
\hline
\end{tabular}
\end{table}
\subsection{Performance Metrics}
Two key metrics, including Root Mean Square Error (RMSE) and 
Average Mean Localization(ALE), which is respectively defined in (14) and (15), is employed here to evaluate the accuracy of the proposed algorithm.
\begin{equation}
\label{eq14}
RMSE=\sqrt{\frac{1}{{{M}_{c}}}\sum\nolimits_{t=1}^{{{M}_{c}}}{\left\| \hat{x}_{u}^{t}-{x}_{u}^{true} \right\|_{2}^{2}}}
\end{equation}
\begin{equation}
\label{eq15}
ALE=\frac{\sum\nolimits_{u=m+1}^{n}{\sqrt{\left\| {{{\hat{x}}}_{u}}-{x}_{u}^{true} \right\|_{2}^{2}}}}{\left( n-m \right)\times R}
\end{equation}
where, ${{M}_{c}},{{\hat{x}}_{u}},x_{u}^{true}$ represent the number of simulation experiments, the estimated position of the unknown node, and the true position of the unknown node, respectively.

\subsection{Results And Analysis}
\subsubsection{Effect of different average hop distances on the proposed algorithm}
The impact of localization accuracy induced by varying average hop distance should to be discussed firstly. This discussion is crucial for understanding how changes in hop distance influence the accuracy and efficiency of node localization in network settings.

\begin{figure}[htbp]
\centering
\includegraphics[width=3.3in]{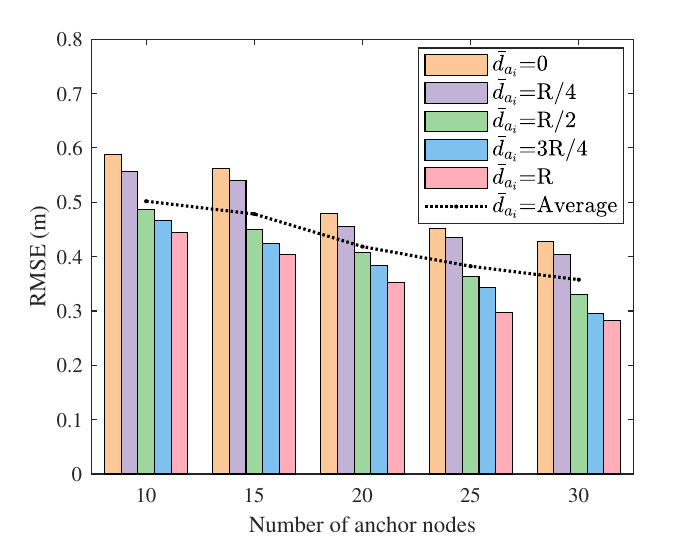}
\caption{Effect of average hop distance on RMSE.}
\label{fig5}
\end{figure}

 Fig.\ref{fig5} displays the RMSE of the proposed AW-MinMax algorithm versus a number of anchors at the various average hop distances. There are two facts to be revealed according to Fig.\ref{fig5}. The first is that the localization accuracy of the AW-MinMax algorithm tends to increase with a number of anchor nodes. It is, of course, due to lots of previous works that have shown that increasing the distribution density of the anchor node may lead to increasing localization accuracy. The second is that the localization accuracy of the AW-MinMax algorithm is also associated with the average hop distance; namely, longer average hop means higher localization accuracy. This conclusion can be demonstrated by the fact that average RMSE values corresponding to different numbers of anchor nodes, represented by dashed lines, tend to be reduced with the number of anchor nodes increasing. Notable, zero average hop distance of anchor node pairs means that those node pairs overlap each other, indicating the number of effective anchor nodes that can be used to localize unknown nodes decreasing. In this case, the localization accuracy of the unknown node tends to be worse (orange color band in Fig.\ref{fig5}).
\subsubsection{ALE comparison of algorithms}
{To thoroughly assess the performance of localization algorithms, simulation experiments are conducted in different scenarios.}

\begin{figure}[htbp]
\centering
\includegraphics[width=3.3in]{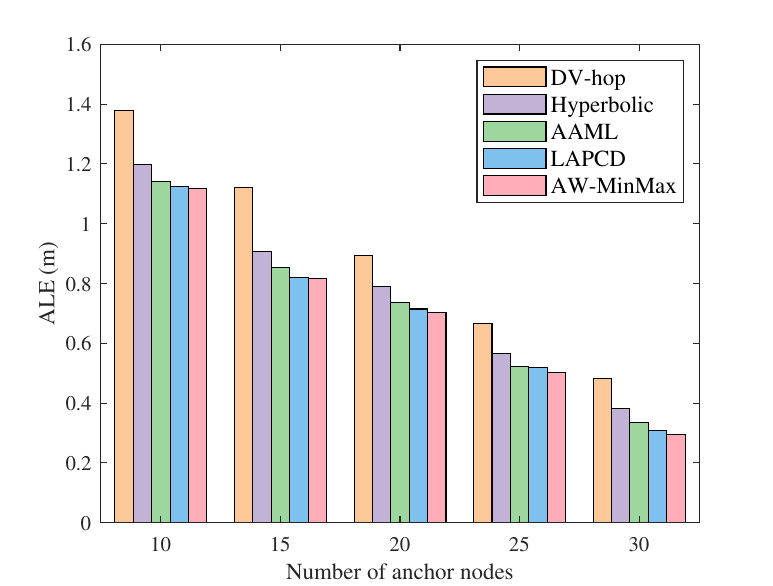}
\caption{The ALE versus anchor node number.}
\label{fig6}
\end{figure}

Fig.\ref{fig6} illustrates the ALE versus anchor node numbers. It is clear that all algorithms tend to achieve higher localization accuracy with the number of anchor nodes increasing, while AW-MinMax experiences the best localization accuracy compared to other algorithms. Specifically, when the number of anchor nodes reaches 30, the proposed AW-MinMax algorithm shows improvements of 38.54\%, 35.82\%, 30.21\%, and 20.90\% compared to DV-Hop, Hyperbolic, AAML, and LAPCD, respectively.
\begin{figure}[htbp]
\centering
\includegraphics[width=3.3in]{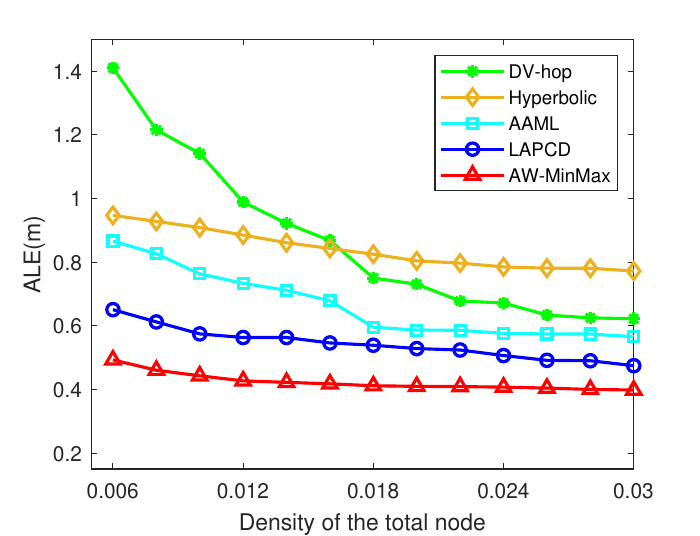}
\caption{The ALE versus node density.}
\label{fig7}
\end{figure}

Additionally, Fig.\ref{fig7} demonstrates the impact of total node density on ALE. At a node density of 0.006, the proposed AW-MinMax algorithm reduces the ALE by 60.05\%, 47.95\%, 43.16\%, and 24.25\% compared to DV-Hop, Hyperbolic, AAML, and LAPCD, respectively. All algorithms exhibit a decreasing trend in ALE as total node density increases. However, due to differences in the models used for node localization, there is a notable intersection between the ALE curves of DV-Hop and Hyperbolic. 



\begin{figure}[htbp]
\centering
\includegraphics[width=3.3in]{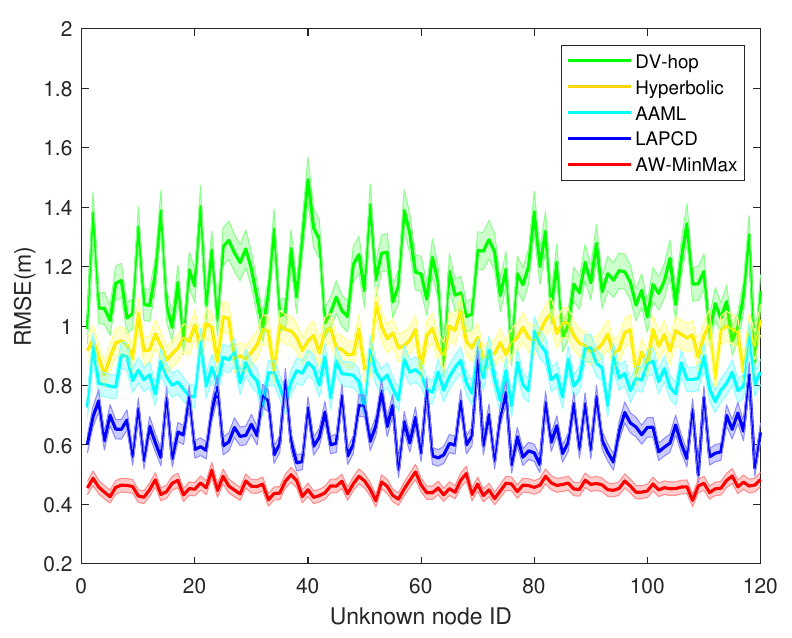}
\caption{The RMSE of unknown node localization.}
\label{fig8}
\end{figure}

\begin{figure}[htbp]
\centering
\includegraphics[width=3.3in]{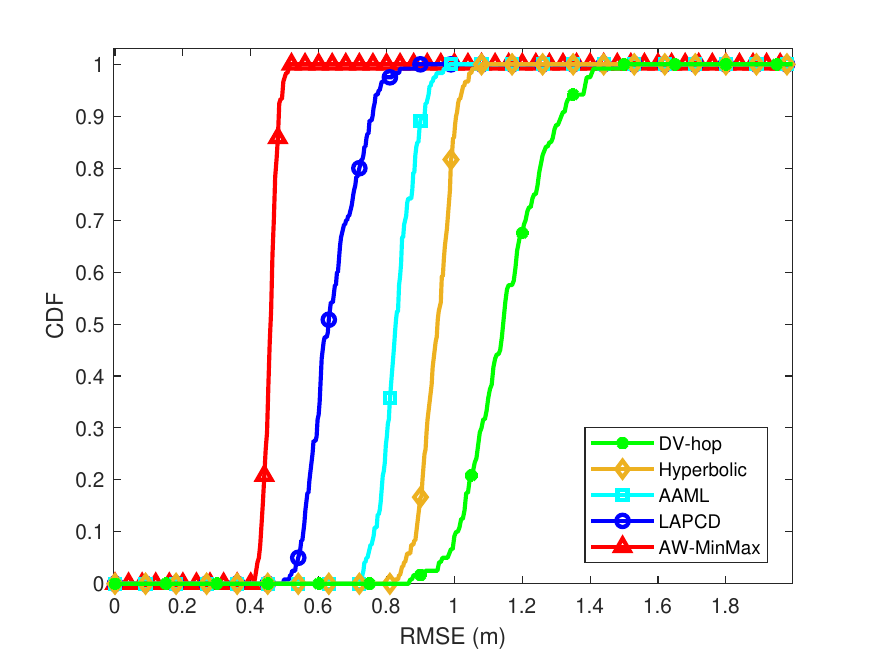}
\caption{The CDF of RMSE.}
\label{fig9}
\end{figure}
By employing the different algorithms, the RMSE of unknown node ID localization is plotted in Fig.\ref{fig8}. This visualization clearly ranks the RMSE of the five algorithms from highest to lowest as follows: DV-Hop, Hyperbolic, AAML, LAPCD, and the proposed algorithm. Upon analysis, it is evident that, compared to the other four algorithms, the proposed AW-MinMax algorithm reduces the average RMSE by 60.25\%, 51.72\%, 45.03\%, and 28.92\%, respectively, indicating that our algorithm achieves the highest localization accuracy.

The cumulative distribution function (CDF) of the localization error of the five algorithms is shown in Fig.\ref{fig9}. The CDF curves of the five algorithms increase monotonically from left to right. Notably, when the CDF reaches 1, the AW-MinMax algorithm shows improvements of 41.89\%, 47.56\%, 52.22\%, and 64.04\% over DV-Hop, Hyperbolic, AAML, and LAPCD, respectively. Additionally, the errors of the proposed algorithm are primarily concentrated between 0.3 and 0.5. In contrast, the ranges for LAPCD, AAML, Hyperbolic, and DV-Hop extend from 0.4 to 0.9, 0.7 to 1, 0.8 to 1.1, and 0.8 to 1.4, respectively. As expected, the proposed algorithm exhibits a more concentrated range of localization errors compared to the others, demonstrating its capability to reduce the peaks of error in node localization within irregular networks. This concentrated error distribution further underscores the enhanced precision and reliability of the AW-MinMax algorithm.

\section{Conclusion}
\label{section6}
This paper introduces the AW-MinMax algorithm, designed to address the range-free node localization problem in irregular networks. By obtaining estimated distances between nodes based on different types of anchor pairs and adaptively assigning weights to unknown node localization estimations using the minimum hop selection criterion, this algorithm enhances the accuracy and stability of localization. Moreover, to minimize the maximum residual of the adaptive weighted unknown node localization estimation, we have formulated an optimization problem that effectively approximates the original range-free node localization challenge. Compared to traditional algorithms, the proposed AW-MinMax algorithm significantly improves localization accuracy and stability, successfully avoiding entrapment in local optima. However, the distance estimation between nodes within the proposed framework relies on multiple complex calculations, which may lead to increased overhead and power costs. Addressing these challenges will be a primary focus for future work to enhance the efficiency and applicability of the AW-MinMax algorithm.

\appendix[A]
In section \ref{section3} A, to obtain ${{p}_{\theta _{{{a}_{i}}}^{u}}}\left( \theta  \right)$, the calculation process is provided in here.
\begin{figure}[htbp]
\centering
\includegraphics[width=2.3in]{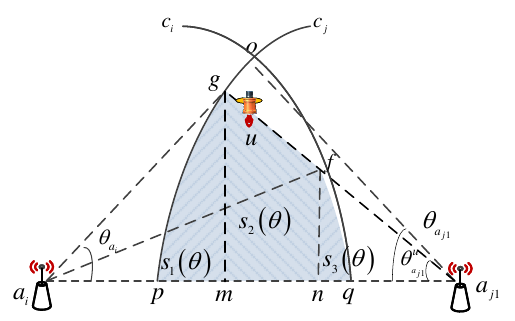}
\caption{Derivation of ${{f}_{\theta _{{{a}_{i}}}^{u}}}\left( \theta  \right)$ in geometric.}
\label{fig10}
\end{figure}
From Fig.\ref{fig10}, the probability of $u$ lies within ${{S}_{pgfq}}$, $\theta _{{{a}_{i}}}^{u}\le \theta $, which equal to ${{f}_{\theta _{{{a}_{i}}}^{u}}}\left( \theta  \right)$:
\begin{equation}
\label{eq16}
{{f}_{\theta _{{{a}_{i}}}^{u}}}\left( \theta  \right)=\frac{{{S}_{pgfq}}}{{{S}_{poq}}}=\frac{{{s}_{1}}\left( \theta  \right)+{{s}_{2}}\left( \theta  \right)+{{s}_{3}}\left( \theta  \right)}{{{S}_{poq}}}
\end{equation}
where ${{s}_{1}}\left( \theta  \right)$,${{s}_{2}}\left( \theta  \right)$,${{s}_{3}}\left( \theta  \right)$ region in ${{S}_{pgfq}}$ are disjoint.

Unlike \cite{b25}, this paper uses the tangent function instead of the cosine function to obtain ${{s}_{1}}\left( \theta  \right)$:

\begin{equation}
\begin{aligned}
\label{eq17}
  {{s}_{1}}\left( \theta  \right) &= {{S}_{{{a}_{j1}}gp}}-{{S}_{{{a}_{j1}}gm}} \\ 
  &= \frac{1}{2}{{R}_{{{a}_{i}}}}^{2}{{\tan }^{-1}}\left[ \frac{\sin \theta \cos \theta ({{d}_{{{a}_{i}}{{a}_{j1}}}}-{{\delta }_{{{a}_{i}}{{a}_{j1}}}})}{{{d}_{{{a}_{i}}{{a}_{j1}}}}-{{\cos }^{2}}\theta {{\delta }_{{{a}_{i}}{{a}_{j1}}}}} \right] \\ 
  &\quad -\frac{1}{4}\sin (2\theta ){{d}_{{{a}_{i}}{{a}_{j1}}}}({{d}_{{{a}_{i}}{{a}_{j1}}}}-{{\delta }_{{{a}_{i}}{{a}_{j1}}}}) \\ 
  &\quad +\frac{1}{4}\sin (2\theta )\cos \theta \left({{d}_{{{a}_{i}}{{a}_{j1}}}}-{{\delta }_{{{a}_{i}}{{a}_{j1}}}}\right)^{2}  
\end{aligned}
\end{equation}
where, ${{R}_{{{a}_{i}}}}$ represents the communication radius of anchor nodes, ${{R}_{{{a}_{i}}}}={{h}_{{{a}_{i}}u}}\cdot R$, ${{\delta }_{{{a}_{i}}{{a}_{j1}}}}$ can be expressed as:
\begin{equation}
\label{eq18}
{{\delta }_{{{a}_{i}}{{a}_{j1}}}}=\sqrt{d_{{{a}_{i}}{{a}_{j1}}}^{2}-(d_{{{a}_{i}}{{a}_{j1}}}^{2}-R_{{{a}_{i}}}^{2}){{\sec }^{2}}\theta }
\end{equation}

Similarly, ${{s}_{3}}\left( \theta  \right)$:
\begin{equation}
\label{eq19}
{{s}_{3}}\left( \theta  \right) ={{S}_{{{a}_{i}}fq}}-{{S}_{{{a}_{i}}fn}} =\frac{1}{2}R_{{{a}_{i}}}^{2}\left( \theta -\cos \theta \sin \theta  \right)
\end{equation}
\begin{equation}
\begin{aligned}
\label{eq20}
 {{s}_{2}}\left( \theta  \right) &={{S}_{{{a}_{i}}gn}}-{{S}_{{{a}_{i}}fm}} \\ 
 & =\frac{1}{2}\left[ {{R}_{{{a}_{j1}}}}\cos \theta -{{\cos }^{2}}\theta ({{d}_{{{a}_{i}}{{a}_{j1}}}}-{{\delta }_{{{a}_{i}}{{a}_{j1}}}}) \right] \\ 
 & \cdot \left[ {{R}_{{{a}_{i}}}}\sin \theta +(\sin \theta \cos \theta )({{d}_{{{a}_{i}}{{a}_{j1}}}}-{{\delta }_{{{a}_{i}}{{a}_{j1}}}}) \right]  
\end{aligned}
\end{equation}

As for ${{S}_{poq}}$:
\begin{equation}
\begin{aligned}
\label{eq21}
  {{S}_{poq}}\left( \theta  \right) &={{S}_{{{a}_{i}}oq}}+{{S}_{{{a}_{j1}}op}}-{{S}_{o{{a}_{i}}{{a}_{j1}}}} \\ 
 & =\frac{1}{2}R_{{{a}_{i}}}^{2}{{\theta }_{o{{a}_{i}}{{a}_{j1}}}}+\frac{1}{2}R_{{{a}_{j1}}}^{2}{{\theta }_{o{{a}_{j1}}{{a}_{i}}}}-{{S}_{o{{a}_{i}}{{a}_{j1}}}} \\ 
\end{aligned}
\end{equation}
where ${{S}_{o{{a}_{i}}{{a}_{j1}}}}$ in (\ref{eq21}), is calculated using the Heron’s equation \cite{b25}:
\begin{equation}
\label{eq22}
{{S}_{o{{a}_{i}}{{a}_{j1}}}}=\sqrt{l\left( l-{{d}_{{{a}_{i}}{{a}_{j1}}}} \right)\left( l-{{R}_{{{a}_{i}}}} \right)\left( l-{{R}_{{{a}_{j1}}}} \right)}
\end{equation}
where $l$ represents half the circumference of the triangle $o{{a}_{i}}{{a}_{j1}}$.

With the above (\ref{eq16})-(\ref{eq21}), ${{p}_{\theta _{{{a}_{i}}}^{u}}}\left( \theta  \right)$ of the (\ref{eq6}) in section \ref{section3} A can be obtained by taking the partial derivative of ${{f}_{\theta _{{{a}_{i}}}^{u}}}\left( \theta  \right)$ with respect to $\theta $:
\begin{equation}
\begin{aligned}
\label{eq23}
  & {{p}_{\theta _{{{a}_{i}}}^{u}}}\left( \theta  \right)=\frac{\partial {{f}_{\theta _{{{a}_{i}}}^{u}}}\left( \theta  \right)}{\partial \theta } \\ 
 &             =\frac{2{{d}_{{{a}_{i}}{{a}_{j1}}}}\left( 2{{R}_{{{a}_{i}}}}{{R}_{{{a}_{j1}}}}+2d_{{{a}_{i}}{{a}_{j1}}}^{2}{{\sin }^{2}}\theta  \right)}{C\sqrt{d_{{{a}_{i}}{{a}_{j1}}}^{2}+{{\sec }^{2}}\theta \left( R_{{{a}_{j1}}}^{2}-d_{{{a}_{i}}{{a}_{j1}}}^{2} \right)}}-\frac{2d_{{{a}_{i}}{{a}_{j1}}}^{2}\cos \left( 2\theta  \right)}{C} \\ 
\end{aligned}
\end{equation}
where $C$ is available from the literature \cite{b25}.

In summary, ${{p}_{\theta _{{{a}_{i}}}^{u}}}\left( \theta  \right)$ of (\ref{eq6}) is solved.


\begin{IEEEbiography}[{\includegraphics[width=1in,height=1.25in,clip,keepaspectratio]{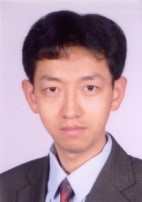}}]{Yong Jin} received B.S. in Electrical Engineering from Tongji University, Shanghai, China, in 1994 and Ph.D. degrees in Information and Communication Engineering from Northwestern Polytechnical University, Xi’an, China, in 2010. Since 2015, he was as an Professor with the College of Computer and Information Engineering, Henan University, Kaifeng, Henan, China. Since 2021, he was as an Professor with the School of Artificial Intelligence, Henan University, Zhengzhou, Henan, China. Also, he has served as a peer-reviewer for various IEEE research journals since 2010. His research interests include array signal processing and statistical signal processing.
\end{IEEEbiography}

\begin{IEEEbiography}[{\includegraphics[width=1in,height=1.25in,clip,keepaspectratio]{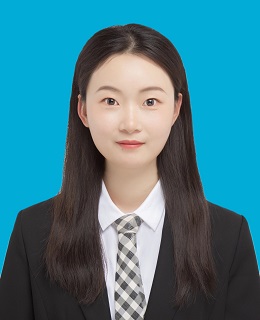}}]{Leng Junfang} received the B.S. degree in electrical engineering and automation from Shangqiu Normal College, China, in 2021, and the M.S. degree in electronic information from Henan University, China, in 2024, and is a student member of intelligent technology and systems laboratory at Henan University. Since 2024, she was as a teacher with the School of Information Engineering, Zhengzhou College of Finance and Economics, Zhengzhou, Henan, China, with a research interest in Wireless Sensor Networks, Positioning and Target Tracking.
\end{IEEEbiography}

\begin{IEEEbiography}[{\includegraphics[width=1in,height=1.25in,clip,keepaspectratio]{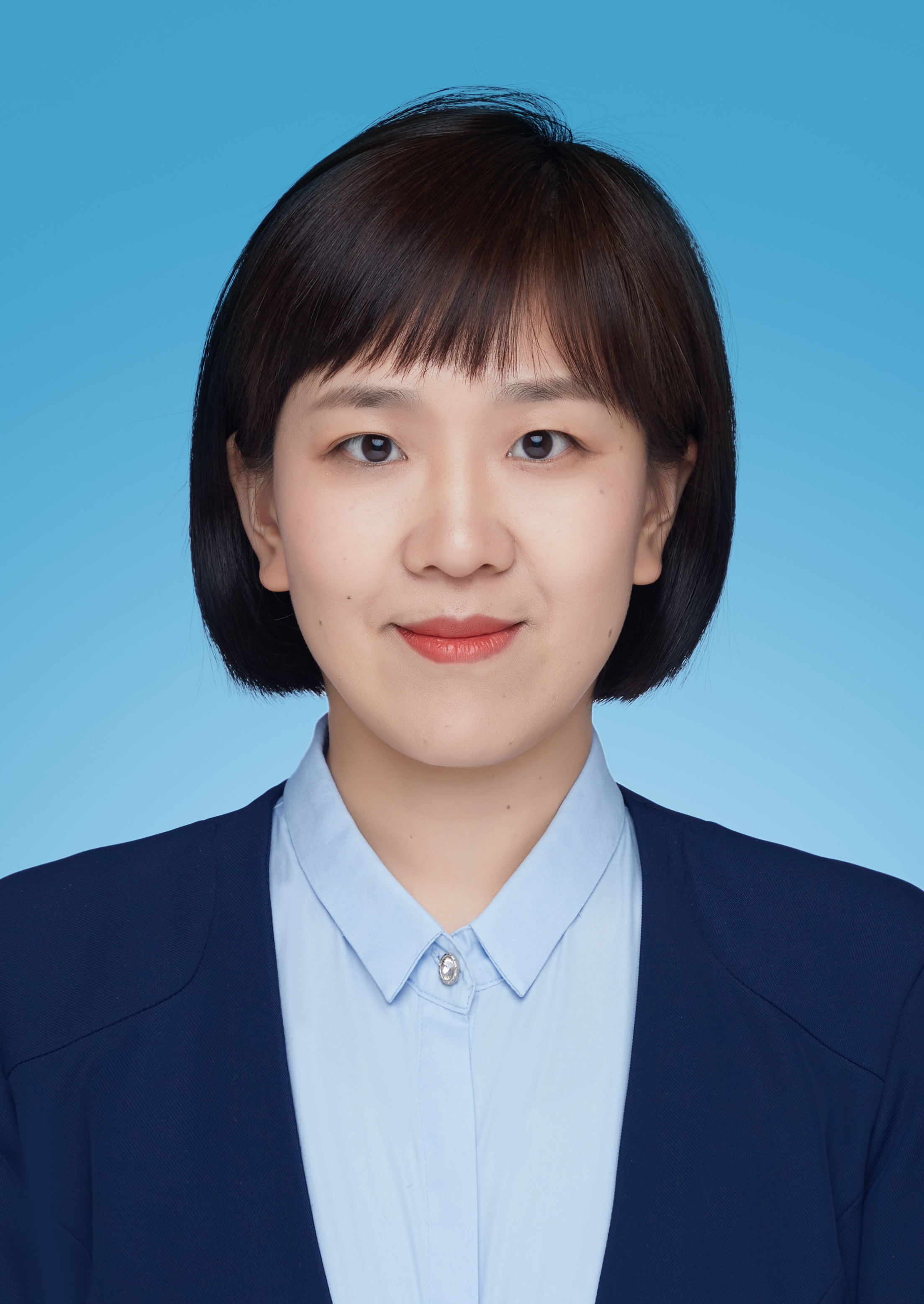}}]{Lin Zhou} received the M.S. in application mathematics from Henan University, China, in 2005, and the Ph.D degree in control theory and control engineering from Northwestern Polytechnical University China, in 2013. Now, she is an Associate Professor of School of Artificial Intelligence, Henan University. Her research interests include information fusion, sensor management.
\end{IEEEbiography}

\begin{IEEEbiography}[{\includegraphics[width=1in,height=1.25in,clip,keepaspectratio]{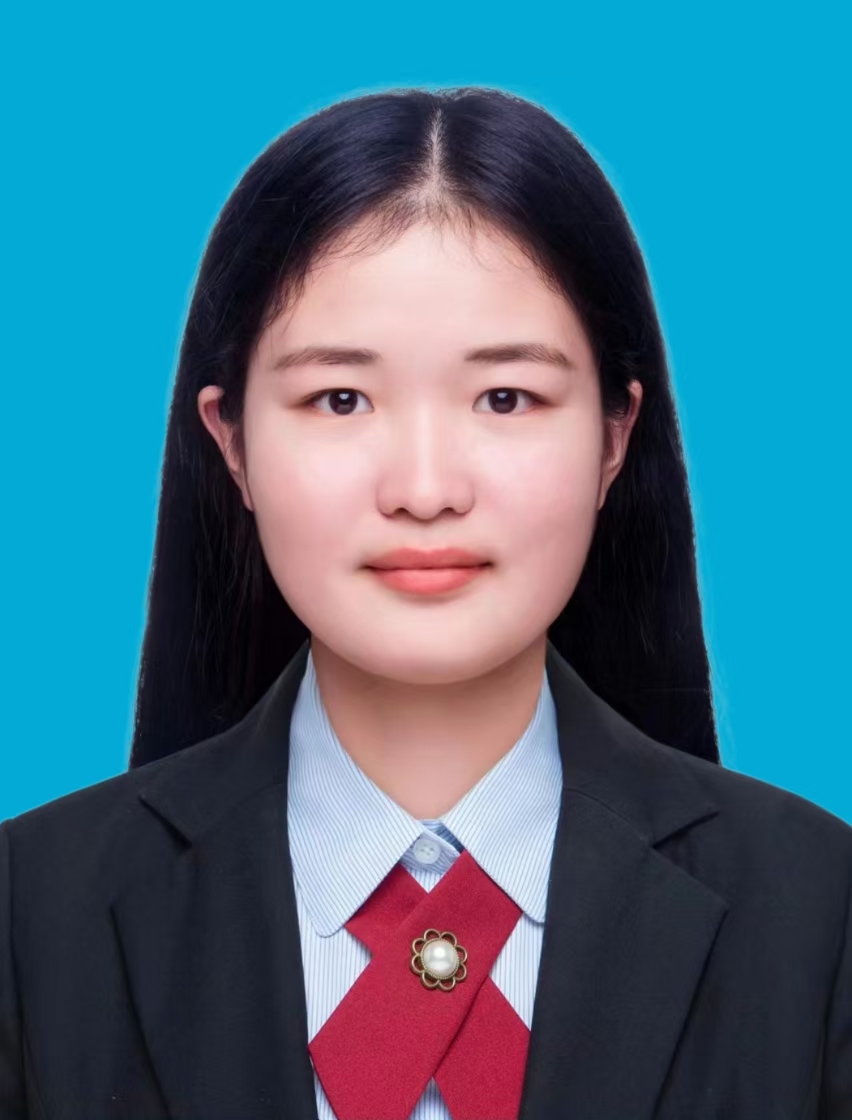}}]{Yu Jiang} received B.S. degree in aircraft quality and reliability from Zhengzhou University of Aeronautics, China, in 2019, and she is currently working toward the M.S. degree in Henan University. Now, she is a student member of the lab of intelligent technology and systems of Henan University and her research interests include wireless sensor networks, localization.
\end{IEEEbiography}

\begin{IEEEbiography}[{\includegraphics[width=1in,height=1.25in,clip,keepaspectratio]{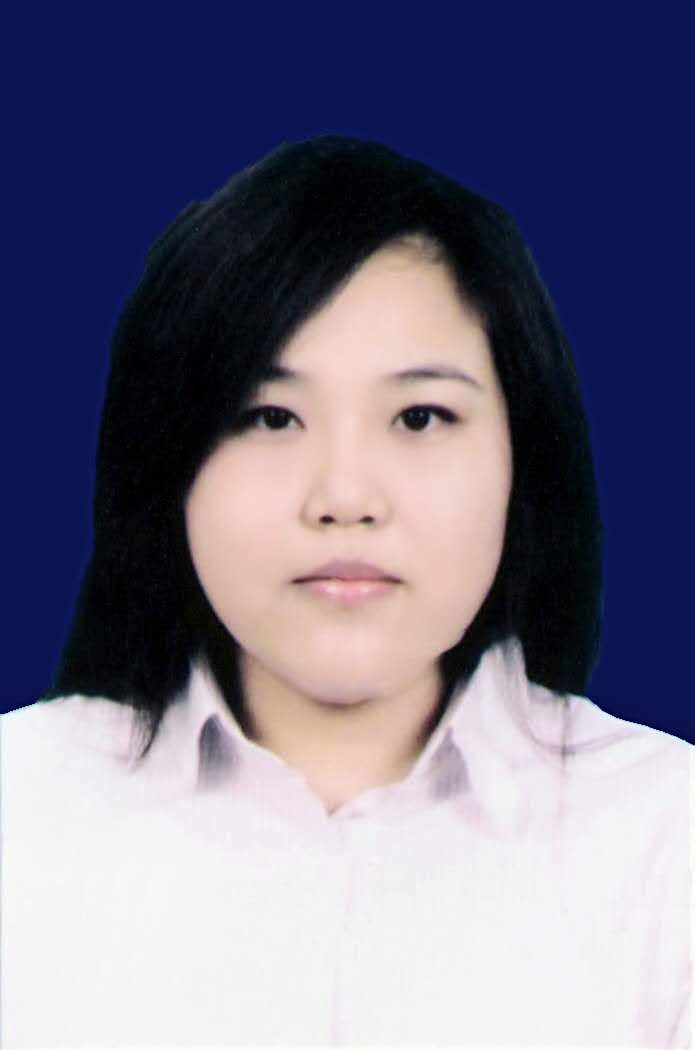}}]{Qian Wei} received the Ph.D degree in control theory and control engineering from Xi'an Jiaotong University, China, in 2016. Now, she is a lecturer of School of Artificial Intelligence, Henan University. Her research interests include aircraft guidance and control, information fusion, sensor management.
\end{IEEEbiography}

\vfill

\end{document}